\documentclass{emulateapj}
\submitted{Accepted for publication in ApJ, 2006 June 28}

\newcommand\psr{AX~J1845.0$-$0258}
\newcommand\cxou{CXOU~J184454.6$-$025653}
\newcommand\axj{AX~J184453$-$025640}
\def\lapp{\ifmmode\stackrel{<}{_{\sim}}\else$\stackrel{<}{_{\sim}}$\fi}
\def\gapp{\ifmmode\stackrel{>}{_{\sim}}\else$\stackrel{>}{_{\sim}}$\fi}

\begin{document}

\title{{\it Chandra} Observations of the Transient 7-s X-ray Pulsar
  \psr}

\author{
C. R. Tam,\altaffilmark{1}
V. M. Kaspi,\altaffilmark{1,2}
B. M. Gaensler,\altaffilmark{3,4}
E. V. Gotthelf,\altaffilmark{5}
}

\altaffiltext{1}{Department of Physics, Rutherford Physics Building,
McGill University, 3600 University Street, Montreal, Quebec,
H3A 2T8, Canada; tamc@physics.mcgill.ca}

\altaffiltext{2}{Canada Research Chair}

\altaffiltext{3}{Harvard-Smithsonian Center for Astrophysics, 60
  Garden St., Cambridge, MA 02138}

\altaffiltext{4}{Alfred P. Sloan Research Fellow}

\altaffiltext{5}{Columbia Astrophysics Laboratory, Columbia
  University, 550 West 120th Street, New York, NY 10027-6601}

\begin{abstract}

We present the results of \textit{Chandra X-ray Observatory}
observations of the transient anomalous X-ray pulsar candidate
\psr\ in apparent quiescence.  Within the source's error circle, we
find a point source and possible counterpart, which we designate 
\cxou.  No coherent pulsations are detected, and no extended emission
is seen.  The source's spectrum is equally well described by a
blackbody model of temperature $kT \sim 2.0$~keV or a power-law model
with photon index $\Gamma \sim 1.0$.  This is considerably harder
than was seen for \psr\ during its period of brightening in 1993 ($kT
\sim 0.6$~keV) despite being at least $\sim$13 times fainter.  This
behaviour is opposite to that observed in the case of the established
transient AXP, XTE~J1810$-$197.  We therefore explore the possibility
that \cxou\ is an unrelated source, and that \psr\ remains undetected
since 1993, with a flux 260$-$430 times fainter than at that epoch.
If so, this would represent an unprecedented range of variability in
AXPs.

\end{abstract}

\keywords{pulsars: general --- pulsars: individual (\psr) --- X-ray:
  individual (\cxou) --- stars: neutron --- stars: pulsars}

\section{Introduction}

The class of neutron stars collectively known as ``Anomalous X-ray
Pulsars'' \citep[AXPs;][]{ms95} has many properties that have been
enigmatic since the discovery of the first example over 20 years ago
\citep{fg81}.  Foremost among puzzles was the nature of their energy
source, as they show no evidence of being either accretion- or
rotation-powered.  Following extensive theoretical and
observational work \citep[see][for a review]{wt06}, it is clear that
AXPs share a common nature with another unusual class of neutron
stars, the ``Soft Gamma Repeaters'' (SGRs), with both best identified
with young, isolated neutron stars that are powered by the decay of an
enormous ($\ga$10$^{14}$~G) internal magnetic field.  As such, they are
called ``magnetars'' \citep{dt92a,td95,td96a}.

Recently, transient X-ray pulsars with properties otherwise unique to
the AXPs have been discovered.  The one established transient AXP
(TAXP) is XTE~J1810$-$197, a 5.5-s X-ray 
pulsar discovered in 2003 \citep{ims+04} during a period of dramatic
X-ray enhancement and subsequent flux decay on roughly a year
timescale. The source's spectrum at the time 
of the outburst was soft in the 2--10~keV band, well characterized by
a combined two-component spectrum (power law plus blackbody, or 2
temperature blackbody model) with parameters similar to those seen in
classical, i.e., non-transient, AXPs \citep{ims+04,ghbb04}.  This,
together with the observed secular spin down and implied
magnetar-strength magnetic field, as well as an observed X-ray
luminosity in excess of the implied rotational spin-down luminosity
$\dot{E}$, makes an AXP interpretation for XTE~J1810$-$197 difficult
to escape \citep{ims+04}.  Yet \citet{ghbb04} showed from archival
X-ray data that in quiescence, the observed source flux was nearly two
orders of magnitude fainter than at the time of the outburst and in
subsequent months, and much fainter than any of the non-transient
AXPs.  TAXPs also open the question of how many more quiescent AXPs
there are in the Galaxy.  This question is particularly
interesting as the magnetar birthrate could be a substantial fraction
of the total neutron star birthrate, possibly even comparable to that
of classical radio pulsars, whose much greater longevity makes them
much more numerous in the Galaxy.  On the other hand, \citet{gmo+05}
consider the growing evidence that magnetars have unusually massive
progenitors, and thus argue that the magnetar birthrate is $\sim$10\%
of the total neutron star birthrate.  The study of TAXPs in
quiescence is important for constraining their true luminosity
function, and hence the size of the Galactic magnetar population.

The 6.97-s X-ray pulsar \psr\ was discovered during a periodicity
search for X-ray sources in the {\it ASCA} archive
\citep{gv98,tkk+98}.  Strong X-ray pulsations having a sinusoidal
pulse profile were seen in data obtained in 1993 from a Galactic Plane
point source that was subsequently shown to be near the center of the
shell supernova remnant G29.6+0.1 \citep{ggv99}.  The long pulse
period and possible association with a young remnant strongly
suggested that \psr\ is an AXP.  Additional support for this
interpretation came from the soft, highly absorbed X-ray spectrum,
which was well described by the Wien tail of a blackbody having $kT
\sim 0.64$~keV, similar to that seen in other AXPs
\citep{gv98,tkk+98}.  The pulsar was not detected in a serendipitous
observation of the region obtained in 1997 as part 
of the \textit{ASCA} Galactic Plane Survey \citep{tkk+98}.
Interestingly, follow-up observations in 1999 revealed the source
\axj\footnote{The updated coordinate of this \textit{ASCA} source,
using the final correction for the systematic coordinate offset
derived in \citet{guf+00}, is $18^{\mathrm{h}}44^{\mathrm{m}}54\fs4,
-02\degr56\arcmin37\farcs7$.} in the original 3$'$ radius 
\textit{ASCA} positional uncertainty region, whose flux was smaller by
a factor of $\sim$10 relative to that of \psr\ in 1993, precluding the
measurement of pulsations or spectral information \citep{vgtg00}.  The
rate of change of the spin frequency has therefore not been measured,
rendering the source as yet unconfirmed as a {\it bona fide} AXP.  It
is plausible that the 1993 observation was obtained shortly after a
major outburst like that seen for XTE~J1810$-$197 and that the source
faded subsequently back to its quiescent level.  X-ray observations in
2001--2003 with \textit{BeppoSAX} MECS, \textit{Chandra} HRC-I and
\textit{XMM-Newton} EPIC reported by \citet{isc+04} revealed a faint
source consistent with \axj\ in position and brightness; a concurrent
attempt to detect its optical/IR counterpart produced an H-band upper
limit of 21 mag.

Here, we report on a series of {\it Chandra X-ray Observatory}
observations of \psr\ in apparent quiescence.  We attempt to re-detect
pulsations and hence constrain the spin-down rate in order to test the
AXP interpretation for this source.  We characterize the behaviour of
this candidate TAXP in a low flux state by determining its spectral
properties, and by searching for low-level flux variability on a
variety of time scales.  Finally, we discuss the likelihood and
implications for a non-detection of the AXP counterpart.

\section{Observations and Analysis}

Seven observations with \textit{Chandra} ACIS-S were obtained between
2003 June~26 and September~14 in timed exposure mode.
Table~\ref{tab:obs} summarizes the observing characteristics.  The
first six were taken in 1/8 subarray mode on the chip ACIS-S3, with a
time resolution of 0.441~s, sufficient to resolve the pulsar signal.
The subarray's small field of view does not cover the full 3$'$ radius
\textit{ASCA} error circle.  Therefore, we used as aim point the
position of the counterpart supplied to us from \textit{Chandra} HRC
observations
($18^{\mathrm{h}}44^{\mathrm{m}}54\fs6, -02\degr56\arcmin53\arcsec$
(J2000); G. Israel, private communication).  The seventh observation
was in ACIS-S full frame mode, for which the time resolution was
3.241~s.  The total exposure length at the above position was
$\sim$80~ks.

Data processing was performed with CIAO~3.2.2 and CALDB~3.0.3
software packages.   We re-performed some steps in the standard
processsing pipeline with updated calibration files, using with the
tool \texttt{acis\_process\_events}. 

\subsection{Imaging}
\label{sec:image}
\begin{figure}
\epsscale{1.0}
\plotone{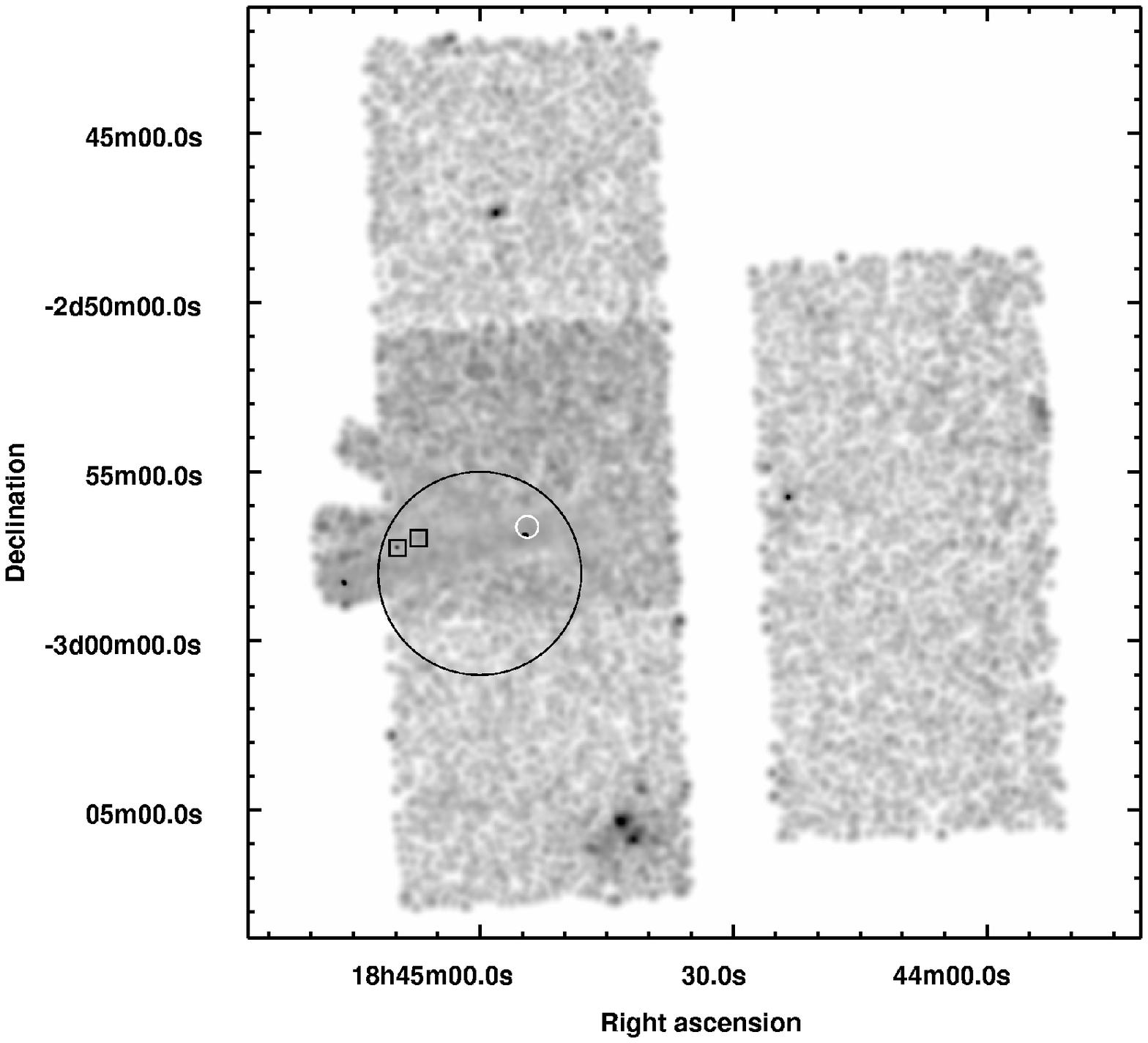}
\figcaption[f1.eps]{Combined \textit{Chandra} ACIS-S image of the
  field surrounding \psr, including chips I2-I3 and S2-S4.  The image
  is  energy-filtered
  (2$-$10~keV), binned by 4 pixels, exposure corrected and
  adaptively smoothed with a (mimimum) 4$''$ Gaussian.  One
  observation (Obs. ID 3896) was omitted due to its effect on the
  appearance of the flat background.  Overlaid are the positional 
  error regions for \psr\ in 1993 \citep[\textit{black
  circle};][]{gv98} and \axj\ in 1999 \citep[\textit{white
  circle}; note that a correction has since been applied to the
  position in][]{vgtg00}.  The point source encircled by both regions
  is \cxou, the possible counterpart to \psr.  Boxes indicate the
  locations of CXOU~J184507.2$-$025657 (\textit{right}; not obvious in
  2$-$10~keV band) and CXOU~J184509.7--025715 (\textit{left}).
\label{fig:image}}
\end{figure}

One source at the position $18^{\mathrm{h}}44^{\mathrm{m}}54\fs68,
-02\degr56\arcmin53\farcs1$ (J2000) is detected in all seven
observations, which we designate \cxou; this is the likely counterpart
to \axj\ and possible counterpart to \psr.  As seen in
Figure~\ref{fig:image}, it falls within the error circles of both
objects.  We find no evidence of extended emission.  The observations
were aligned and summed using the nominal {\it Chandra} astrometric
information.  Systematic uncertainties in {\it Chandra} absolute
positions are expected to be $0.6''$ at the 90\%
level\footnote{http://cxc.harvard.edu/cal/ASPECT/celmon}.  Although
these systematic errors in general can be reduced by aligning other
sources, given that some of our subarray fields contain none, the
nominal astrometry must suffice.  To confirm that co-addition had no
adverse effects on our source's radial profile, we directly compared
it to the simulated PSF produced by the Chandra Ray Tracer (ChaRT) at
the source chip position and found it consistent with an unresolved
point source.  The final position was determined from the combined
image and is consistent with that measured with \textit{Chandra} HRC.

Since the absence of pulsations precludes confirming the AXP nature of
\cxou\ (see \S\ref{sec:time}), and similarly \axj, the true
counterpart could conceivably lie anywhere in the original 3$'$ radius
\textit{ASCA} error circle.  We searched the combined event file,
unfiltered in energy, for additional significant point sources using
\texttt{celldetect}.  One source, CXOU~J184507.2--025657, was found at
$18^{\mathrm{h}}45^{\mathrm{m}}07\fs27, -02\degr56\arcmin57\farcs3$,
located 3$\farcm$1 away from \cxou, at the 3$\sigma$ level.  Its
coincidence with the near-IR source
2MASS~J18450724--0256571\footnote{See
http://www.ipac.caltech.edu/2mass/ for information on the 2MASS All 
Sky Survey} of magnitude $K=12.7$ may suggest that
CXOU~J184507.2--025657 is an unlikely counterpart to \psr, since a 
highly absorbed AXP candidate is expected to have a near-IR magnitude
$K \gg 20$ \citep[for a summary of AXP IR magnitudes and X-ray 
absorptions, see][]{dv05}.  A considerably fainter source,
CXOU~J184509.7--025715 located at
$18^{\mathrm{h}}45^{\mathrm{m}}09\fs76, -02\degr57\arcmin15\farcs0$,
was found at the 2$\sigma$ level.  All of these sources are indicated
in Figure~\ref{fig:image}.  We also inspected an archival
\textit{XMM-Newton} observation, taken 2003 March~3 \citep{isc+04},
but found no additional significant point sources in the error
region.

\subsection{Timing}
\label{sec:time}
In an attempt to perform phase-coherent timing, we observed in 1/8
subarray mode to acquire high-time-resolution data and identify a
pulsed signal.  Light curves for \cxou\ were 
extracted from a $2\farcs5$ radius circle in each data set at the 
maximum allowable time resolution (0.441 s for observations
3891$-$3896) in 3 energy ranges:  1$-$10~keV, 1$-$3~keV and
3$-$10~keV.  Event times were corrected to solar system
barycenter arrival times.  We performed a fast fourier transform (FFT)
on each data set; no evidence for pulsations was found in the
resulting power density spectra.  Using the longest of the
observations (Obs. ID 3891), for the frequency range 0.0880--0.1436~Hz,
we set a 95\% confidence upper limit on the pulsed amplitude of 80\%
in 1$-$10~keV, using the method outlined in \citet{vvw+94}.  The above
frequency range allows for a 10-year change in frequency corresponding
to magnetic fields $\sim$10$^{16}$~G and lower.

Our detections of the faint point sources CXOU~J184507.2--025657 and
CXOU~J184509.7--025715 have far too few counts ($\leq$12 in 1$-$10
keV) to make detecting pulsations possible.

\subsection{Spectrum}
We used the \texttt{psextract} script and \texttt{mkacisrmf} tool to
extract \cxou's spectra from a $2\farcs5$ radius circle and the
background spectra from a 3$''$ to 22$''$ annulus centered on the
point source, and compute instrumental response files.
Background-subtracted count rates for the point source at each 
observing epoch are given in Table~\ref{tab:obs}, where uncertainties
assume Poisson statistics.  At each epoch there were too few counts to
allow a meaningful spectral fit; therefore, we combined the individual
data sets into a summed spectrum containing $550\pm24$
background-subtracted counts (0.5$-$10~keV).  We excluded channels at
energies below 0.5~keV, where the effective area of ACIS-S falls off
significantly, and grouped the remainder so that a minimum of 12
counts fell in each spectral bin.  The spectral fitting package
XSPEC~11.3.1 produced equally acceptable fits to single-component
thermal blackbody or power-law models with photoelectric absorption;
model parameters are presented in Table~\ref{tab:spec}.  We found a
best-fit temperature of $kT = 2.0^{+0.4}_{-0.3}$~keV and an absorption
of $N_H=5.6^{+1.6}_{-1.2} \times 10^{22}$~cm$^{-2}$ assuming a
blackbody spectrum, and a photon index of $\Gamma = 1.0^{+0.5}_{-0.3}$
and absorption $N_H=7.8^{+2.3}_{-1.8} \times 10^{22}$~cm$^{-2}$
assuming a power-law spectrum (uncertainties reflect 90\% confidence).
The measured absorptions are consistent with the 1993 \textit{ASCA}
values within uncertainties.

Assuming that the shape of the combined spectrum is also
characteristic for the spectra at each epoch, we determined those
fluxes by holding the spectral parameters fixed at the values in 
Table~\ref{tab:spec}.  Since neither model is preferred based on
goodness of fit, we arbitrarily chose the blackbody model for the rest
of our analysis.  We measured the 2$-$10~keV flux of the seven
individual data sets by grouping spectra in the same way as for the
combined spectrum, freezing $N_H$ and $kT$ at the above best-fit
values, and allowing only the normalization to vary.  We found that
the data are consistent with the source's flux being stable over the
12-week observing window to within statistical uncertainties: fitting
to a constant flux resulted in a reduced $\chi^2=1.0$ for 6 degrees of
freedom.  The inset plot of Figure~\ref{fig:flux} shows the
\textit{Chandra} fluxes assuming the best-fit blackbody model
parameters.

The combined 2$-$10~keV flux of \cxou, assuming the blackbody
spectrum, is $2.6 \pm 0.2 \times 10^{-13}$~erg~s$^{-1}$~cm$^{-2}$.
Although subject to large uncertainties on the measurement of $N_H$,
we estimate that the unabsorbed flux is $2.5-4.0\times
10^{-13}$~erg~s$^{-1}$~cm$^{-2}$.  See Table~\ref{tab:spec} for an
explanation of the uncertainties.  If the source is indeed the
counterpart to \psr, this implies that the flux in 2003 is a factor of
$\sim$13 fainter than that measured in 1993 with \textit{ASCA} GIS
\citep{gv98}.


What if \cxou\ is unrelated to \psr?  We next looked at the two
fainter point sources coincident with the error region as possible
counterparts.  We extracted spectra for CXOU~J184507.2$-$025657 from a
4$''$ radius circle, using the same background area as earlier.  This
source, which was visible in only 3 of the 7 observations, produced
37 background-subtracted counts (0.5$-$10~keV) in its combined
spectrum, insufficent to adequately fit a spectral model.  However, we
observed that the majority of counts fell below 2~keV, contrary to
what one would expect from a highly-absorbed source such as \psr\ that
previously exhibited $N_H \ga 6 \times 10^{22}$~cm$^{-2}$.  This
evidence, combined with the probable 2MASS association we mentioned
earlier, strongly suggests that CXOU~J184507.2$-$025657 is unrelated
to \psr.  From CXOU~J184509.7--025715, which appeared in 4 of 7
observations, we extracted counts from a 4$\farcs$4 radius circle;
this gave a combined spectrum containing 20 background-subtracted
counts (0.5$-$10~keV).  Again, the paucity of counts prevented us from
drawing any conclusive results about the spectrum of this source.

Finally, we considered the case that \psr\ was not at all redetected,
and determined the $3\sigma$ upper limit on the absorbed flux for a 
hypothetical point source.  We measured the background count rate from
our only full-frame data set (Obs. ID 3897), which was the only
observation whose field was large enough to contain the full 3$'$
\textit{ASCA} error circle.  The range $\sim8-13\times 
10^{-15}$~erg~s$^{-1}$~cm$^{-2}$ (2--10~keV) encompasses results
assuming several likely models based on the outburst spectrum of \psr\
and the spectrum of XTE~J1810$-$197 in quiescence (see
Figure~\ref{fig:flux}).  If the true counterpart were off-axis by
3$'$, the difference in effective area and PSF would not dramatically
affect our ability to detect a point source unless it were at the
limiting flux.

\section{Discussion}
\label{sec:disc} 
\begin{figure}
\epsscale{0.8}
\plotone{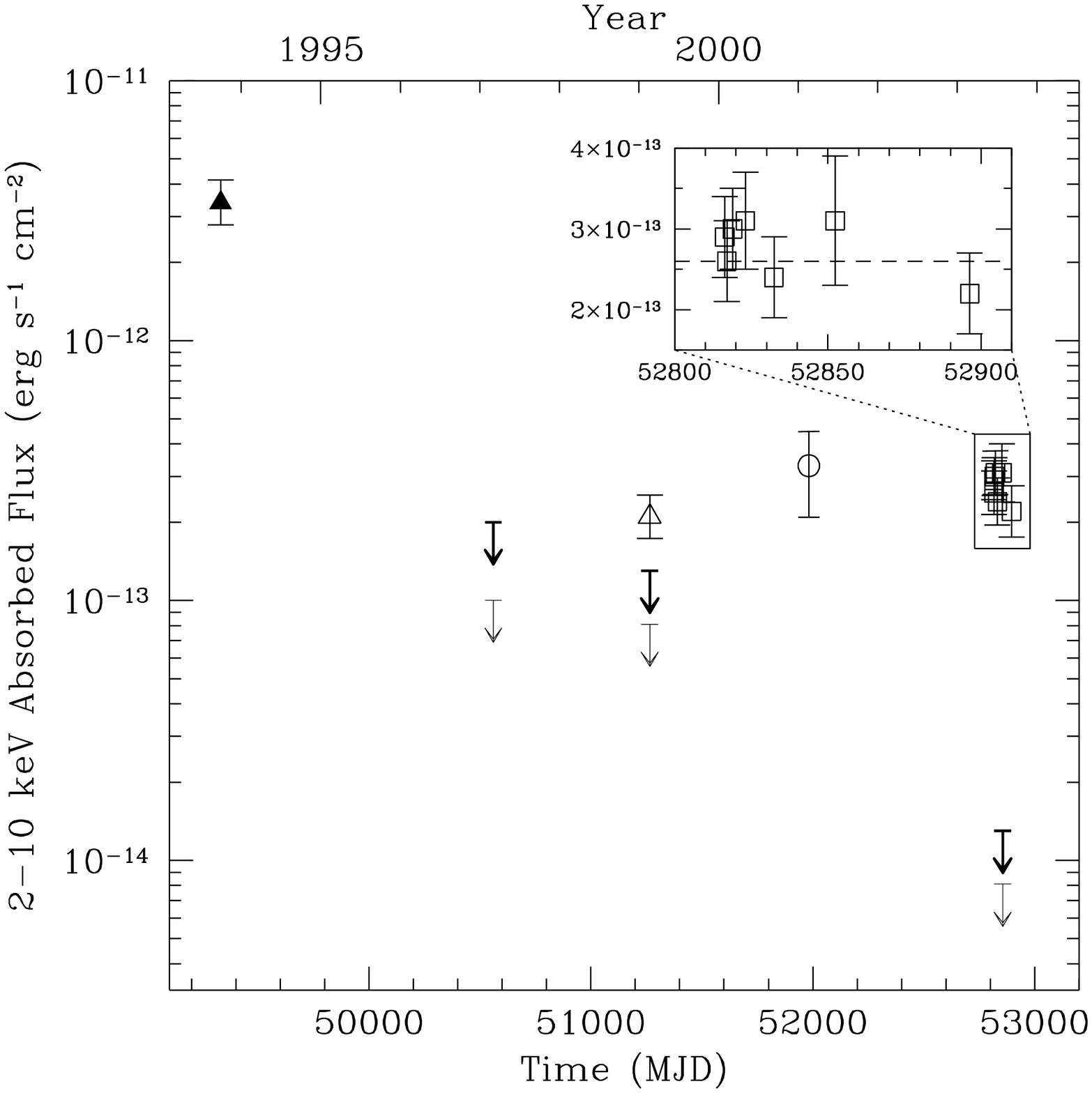}
\figcaption[f2.eps]{The absorbed 2--10~keV flux history of \psr\
  spanning 10~years, given several likely spectral models.
  For the original \textit{ASCA} discovery \citep[\textit{filled
  triangle};][]{gv98}, during which
  pulsations were detected, and the subsequent detection of the faint
  source and
  possible counterpart \axj\ \citep[\textit{open triangle};][]{vgtg00},
  from which neither pulsations nor a spectrum were seen, we have
  adopted the
  outburst blackbody spectrum of \psr\ ($kT\sim 0.64$~keV).
  We show the blackbody flux of a possible counterpart observed with
  \textit{BeppoSAX} (\textit{circle}) as reported by \citet{isc+04}. 
  The flux of \cxou\ (\textit{squares}) assumes the best-fit blackbody
  spectrum described in the text 
  and Table~\ref{tab:spec}.  At some epochs, we plot upper limits in
  addition to detected values, in case \psr's true counterpart fell
  below sensitivity.  Two models are assumed in
  estimating the upper
  limit fluxes: the spectrum of \psr\ in outburst (\textit{thick
  arrows}), and the spectrum of XTE~J1810$-$197 during quiescence
  ($kT\sim 0.18$~keV; \textit{thin arrows}).  The 1997 upper
  limits were measured in an observation of the \textit{ASCA} Galactic
  Plane Survey \citep{tkk+98}. The inset plot
  shows an enlargement of the seven \textit{Chandra} detections, and
  the flux derived from the combined data (\textit{dashed line}).
\label{fig:flux}}
\end{figure}

Our observations reveal that \cxou, whether the counterpart or not, is
significantly fainter than was \psr\ in 1993 \citep{gv98} by a factor
of $\sim$13.  If \cxou\ is not the AXP counterpart, this factor
increases significantly:  \psr\ must now be at least 260--430 times
fainter than it was in 1993.  This would be an unprecedented 
range of variability in AXPs.  \cxou's flux is consistent with that of
\axj\ in 1999 \citep{vgtg00}; therefore, we may well have detected the
same source.  Figure~\ref{fig:flux} summarizes the flux history of
\psr.  

Such variability on long time scales, seen here and in
XTE~J1810$-$197, presents a challenge to the magnetar model, which
posits that the decay of the internal field is continual during the
source's youth.  This decay results in continual internal heating and
crustal stresses.  Thus, the behaviour exhibited by TAXPs raises the
following important question:  if they are magnetars, what causes
the dramatic difference in intrinsic brightness between active and
quiescent states?  Estimates of the crustal temperatures heated by 
internal magnetic dissipation predict X-ray luminosities like those
observed for non-transient AXPs \citep{td96a}.  Those same estimates
are consistent with the expected stresses that result in the crustal
yields that produce bursts like that seen in XTE~J1810$-$197
\citep{wkg+05} and also in the non-transient AXP 1E~2259+586
\citep{kgw+03}.  Thus TAXPs and non-transients have much in common
physically, but are apparently sufficiently dissimilar that their
quiescent X-ray luminosities differ by orders of magnitude.

The spectrum of \cxou\ raises doubt that this is indeed the pulsar
counterpart.  For the blackbody model, the temperature of 2~keV is
much higher than the 0.18~keV measured for XTE~J1810$-$197 in
quiescence \citep{ghbb04} and in fact much higher than for any known 
AXP or SGR\footnote{See online magnetar catalog
  http://www.physics.mcgill.ca/$\sim$pulsar/magnetar/main.html for a 
summary of AXP properties.}.  Evidence for inconsistent spectral
behaviour may have already been seen in 2001--2003 by
\citet{isc+04}. Indeed the {\it Chandra} point source is much 
harder than was the pulsar when in 
outburst in 1993 \citep{gv98,tkk+98}, in stark contrast to
XTE~J1810$-$197 which greatly hardened ($kT=0.67$~keV) when bright.
Thus if \cxou\ is the pulsar counterpart, its spectral properties in
quiescence are puzzling.  The quiescent spectrum is more in line with
that seen from magnetospheric emission in rotation-powered pulsars
\citep[see][for a review]{krh06}, however, no such object has ever
shown even a small variation in its X-ray luminosity, much less orders
of magnitude.  Moreover, the 7-s periodicity is much longer than has
been seen in any rotation-powered magnetospheric X-ray
emission.  The measured 80\% pulsed fraction is well above that
seen in other AXPs, and therefore unconstraining.

If the {\it Chandra} source is not the pulsar counterpart, what could
it be?  The source's salient properties are its hard spectrum, its
approximate luminosity ($L_x \simeq 10^{33}(d/5 \; {\rm kpc})^2$), and
its absence of variability on time scales of days to weeks.  Given the
photon index in the power-law spectral model, an active galactic
nuclei (AGN) interpretation is plausible
\citep[e.g.][]{woau04,nla+05}.  We estimate the 
probability of our object being a background AGN from the predicted
number density as a function of 2$-$10~keV flux according to
\textit{Chandra} ACIS-I deep observations of an ``empty'' Galactic
plane region by \citet{etp+05}.  Coincidentally, their field of view
is centered only $\sim$1$\degr$ from our target, so it is likely
that our fields share many common properties, such as absorption 
column.  From Figure~24 of \citet{etp+05}, the number of extragalactic
point sources per square degree with flux greater than $3\times
10^{-13}$~erg~s$^{-1}$~cm$^{-2}$ is $\sim$2.  We expect $\sim$0.02 AGN
per circular region of radius 3$'$; hence, there is a $\sim$2\% chance
that this source is an AGN.  
\citet{etp+05} estimate $JHK_S$ magnitudes of 21$-$23~mag for AGN in
their survey of the Galactic plane, where at least $A_{K_S}\sim 4$~mag
of extinction may be present \citep{ps95,rl85}.  
Even if we make the simple assumption that the X-ray and near-IR
emission of AGN are part of a single power law spectrum, and that
\cxou\ should be 1$-$2 orders of magnitude brighter in the near-IR
than their survey sample as it is in X-rays, it will still be
difficult to confirm or rule out this possibility using IR
observations, given its predicted faintness.

On the other hand, several types of Galactic objects could have
properties similar to those of this source \citep[see][for a similar 
discussion]{mab+04}. Winds from massive stars have similar spectral
and flux properties, as do some high-mass X-ray binaries.  However,
these would tend to be IR-bright, in conflict with the $H$-band limit
of 21 mag reported by \citet{isc+04}.  A very small number of
millisecond pulsars with similar X-ray luminosities are shown to
possess comparably hard X-ray spectra \citep{kh04}, although until
pulsations are seen from \cxou, it will be impossible to test this.
One source class whose
properties are similar to that of the point source in question are
cataclysmic variables, specifically the class of intermediate polars.
The observed near-IR emission is thought to be dominated 
by their dwarf companion and may be very faint given the
absorption to this source; \citet{mab+04} estimate $K\approx
22-25$~mag for sources at the Galactic center, at comparable distance
and suffering comparable extinction.  This would be hard to detect.

Thus it seems clear that simply obtaining deeper near-IR observations
will not be sufficient to determine whether this source is the
counterpart.  The most promising avenues for doing so therefore are
either obtaining very deep X-ray observations in the hope of
redetecting pulsations, or else waiting patiently for the pulsar to
grace us with another outburst bright enough for follow-up with other
observatories.

We conclude that no matter what, \psr\ is interesting:
if the counterpart is the detected source, then either \psr\ is
not an AXP or AXPs can have a much wider range of spectral properties
in quiescence than has been thought.  If this is not the counterpart
and the AXP
identification is correct, then AXPs are capable of $>$2
order-of-magnitude flux variations, an interesting challenge to the
magnetar model, and also further evidence for a large as-yet-undetected
population.

\acknowledgements
We thank G. Israel for providing the HRC position and
\textit{BeppoSAX} flux and uncertainties of the possible 
counterpart to \psr.  V.M.K. acknowledges funding from NSERC via a
Discovery Grant and Steacie Supplement, the FQRNT, and CIAR.  B.M.G
acknowledges support from Chandra GO grant GO3-4089X, awarded by the
SAO.

\bibliographystyle{apj}
\bibliography{myrefs,journals1,vk_modrefs,psrrefs,crossrefs}

\newpage

\begin{deluxetable}{cccccc}
\tabletypesize{\footnotesize}
\tablewidth{0pt}
\tablecaption{\textit{Chandra} observing characteristics \label{tab:obs}}
\tablehead{
\colhead{Observation ID} & \colhead{Start Date} & \colhead{Start time}
& \colhead{Exposure length} & \colhead{Frame time} & \colhead{Count
rate\tablenotemark{a}}\\
\colhead{} & \colhead{(MJD)} & \colhead{(UT)} & \colhead{(s)} &
\colhead{(s)} & \colhead{(10$^{-3}$ cts s$^{-1}$)} }
\startdata
3891 & 52816.294593424 & 2003-06-26 07:04:12 & 17497 & 0.441 & $6.6 \pm 0.6$\\
3892 & 52817.011209629 & 2003-06-27 00:16:08 & 11011 & 0.441 & $6.3 \pm 0.8$\\
3893 & 52818.904481989 & 2003-06-28 21:42:27 & 11661 & 0.441 & $7.2 \pm 0.8$\\
3894 & 52823.081440211 & 2003-07-03 01:57:16 & 11018 & 0.441 & $7.6 \pm 0.8$\\
3895 & 52832.407594110 & 2003-07-12 09:46:56 & 10930 & 0.441 & $5.9 \pm 0.7$\\
3896 & 52852.389608127 & 2003-08-01 09:21:02 & \phantom{1}7030 & 0.441 & $7.9 \pm 1.1$\\
3897 & 52896.325185093 & 2003-09-14 07:48:15 & 11779 & 3.241 &  $5.9 \pm 0.7$\\
\enddata
\tablenotetext{a}{\cxou\ background subtracted count rates for
  2$-$10~keV energy range.  Errors reflect 1$\sigma$ uncertainties
  assuming Poisson statistics.}
\end{deluxetable}

\newpage

\begin{deluxetable}{lccccc}
\tabletypesize{\footnotesize}
\tablewidth{0pt}
\tablecaption{Spectral properties of combined data \label{tab:spec}}
\tablehead{
\colhead{Model} & \colhead{$N_H$} & \colhead{$kT$ (keV) or $\Gamma$} &
\colhead{$f_{abs}$} & \colhead{$\chi^2$/dof} & \colhead{$f_{unabs}$}\\
\colhead{} & \colhead{(10$^{22}$ cm$^{-2}$)} & \colhead{} &
\colhead{(10$^{-13}$ erg/s/cm$^2$)} & \colhead{} & \colhead{(10$^{-13}$
  erg/s/cm$^2$)} }
\startdata
BB & $5.6^{+1.6}_{-1.2}$ & $2.0^{+0.4}_{-0.3}$ & $2.6\pm0.2$ & 39.9/40 & $2.5-4.0$\\
PL & $7.8^{+2.3}_{-1.8}$ & $1.0^{+0.5}_{-0.3}$ & $2.8\pm0.2$ & 39.9/40 & $2.9-5.0$\\
\enddata
\tablecomments{All errors reflect 90\% confidence intervals.  Absorbed
  and unabsorbed fluxes are given for 2$-$10~keV energy range.
  Uncertainties on absorbed flux reflect the fractional error on the
  normalization assuming the best-fit $N_H$ and $kT$ or $\Gamma$. 
  Unabsorbed flux ranges are found by fixing spectral
  parameters at their 90\% confidence boundaries.}
\end{deluxetable}

\newpage

\newpage

\end{document}